# Playing the Blame Game with Robots[*]

Markus Kneer[†]
Department of Philosophy
University of Zurich
Zürich, Switzerland
markus.kneer@uzh.ch

Michael T. Stuart
Carl Friedrich von Weizsäcker-Zentrum
University of Tübingen
Tübingen, Germany
mike.stuart.post@gmail.com

## ABSTRACT

Recent research shows – somewhat astonishingly – that people are willing to ascribe moral blame to AI-driven systems when they cause harm [1]–[4]. In this paper, we explore the moral-psychological underpinnings of these findings. Our hypothesis was that the reason why people ascribe moral blame to AI systems is that they consider them capable of entertaining inculpating mental states (what is called *mens rea* in the law). To explore this hypothesis, we created a scenario in which an AI system runs a risk of poisoning people by using a novel type of fertilizer. Manipulating the computational (or quasi-cognitive) abilities of the AI system in a between-subjects design, we tested whether people's willingness to ascribe knowledge of a substantial risk of harm (i.e., recklessness) and blame to the AI system. Furthermore, we investigated whether the ascription of recklessness and blame to the AI system would influence the perceived blameworthiness of the system's *user* (or *owner*). In an experiment with 347 participants, we found (i) that people are willing to ascribe blame to AI systems in contexts of recklessness, (ii) that blame ascriptions depend strongly on the willingness to attribute recklessness and (iii) that the latter, in turn, depends on the perceived "cognitive" capacities of the system. Furthermore, our results suggest (iv) that the higher the computational sophistication of the AI system, the more blame is shifted from the human user to the AI system.

## CCS CONCEPTS

• **Computing methodologies ~ Artificial intelligence ~ Philosophical/theoretical foundations of artificial intelligence ~ Theory of mind • Computing methodologies ~ Artificial intelligence ~ Knowledge representation and reasoning ~ Reasoning about belief and knowledge** • Human-centered computing ~ Human computer interaction (HCI) ~ Empirical studies in HCI

## 1 Introduction

Philosophers and computer scientists have repeatedly cautioned against adopting psychological language towards artificially intelligent systems, as this can lead to "premature conclusions of ethical or legal significance" [5, p. 166-7] [5]–[7]. Differently put, since (on most views) moral agency requires the capacity for inculpating mental states, postulating the latter for AI systems might engender the mistaken inference that they can be moral (and legal) agents [8], [9].

Nevertheless, psychological and sociological investigations show that people are willing to attribute rich mental states to currently existing AI systems [1], including foreknowledge of bad outcomes [10] and intentions to deceive [11]. They are also willing to treat such systems as blameworthy [2]–[4], [12], [13]. In some studies, participants see AI systems as being less blameworthy than humans, for example, when comparing human and AI-driven cars that strike a pedestrian [12]. But even here, almost half of the participants see the autopilot as blameworthy. Meanwhile, other studies have shown that, in certain situations, people attribute *more* blame to AI systems than to human agents when everything apart from agent type is held fixed [2], [10], [13].

Why are people making these judgments? As Malle and colleagues have shown, anthropomorphic AI systems are treated more like humans than their mechanical-looking counterparts, as far as morality is concerned ([14], see also instanceproject.eu). Perhaps people are more willing to blame anthropomorphic AI systems because looking human naturally leads people to ascribe mental traits [1], [8]. Going beyond inferences based on the physical appearance of the robot, we decided to target the connection between the perceived *capacity* for inculpating mental states and moral evaluations directly. In a previous experiment on this relationship, we found that even when people were given the option to downgrade their attributions of inculpating mental states from literal to metaphorical (e.g., people had the choice between classifying an AI system as knowing what it was doing versus merely "knowing" what it was doing – in scare quotes), participants mostly refused. Participants attributed mental states to AI systems to the same degree as to human agents, or group agents (corporations) [10]. Another study found that people are just as willing to attribute intentions to deceive and lying behavior to AI systems as to human agents [11].

Again, why might this be? Perhaps participants attribute inculpating mental states to AI systems because in extant studies (ours included), the systems and their actions are described in ways that suggest something close to human agency (and the capacities the latter entails). In the experiment reported below, which explores the relation between perceived mental properties and perceived moral capacity, we controlled for this by describing AI systems that differed in terms of computational sophistication. Our question was: as computational sophistication increases, are people more willing to ascribe the mental capacities sufficient for perceived moral agency, and hence blame them more? And further: is there a minimal set of properties that the system requires for people to deem it sufficiently agent-like, and thus amenable to the attribution of moral blame?

Our investigation focused on *epistemic* mental states, specifically, the *knowledge* an agent has of a potential risk. If an agent has knowledge of a risk, and enacts risky behavior anyway, this qualifies the agent as acting recklessly (see Model Penal Code 2.0.2.c). Whether an entity can be deemed reckless depends strongly on the kinds of capacity it has. It makes little sense, for instance, to say of a toaster (a simple machine), that it should have known better than to continue operating when doing so would result in a fire. It is also doubtful whether toddlers (human agents who are not yet capable of full responsibility) can be considered reckless. However, we *can* consider an adolescent who is playing with matches reckless. At what point, then, are we willing to ascribe recklessness – and thus a core requisite for minimal moral agency – to an AI system?[1] And to what extent are people willing to transfer blame for bad actions to human agents who stand in some suitable relation with an AI system (e.g., owning it, or using it), when the latter risks some serious harm?

## 2 Experiment

### 2.1 Participants

We recruited 374 participants on Amazon Mechanical Turk. IP addresses were restricted to the United States. Those who failed an attention test or a comprehension check were excluded, leaving 347 participants (age M=42 years, SD=13 years; 168 females).

### 2.2 Participants, Methods, and Materials

In the scenario (see Appendix), Shill & Co., a farming company, relies on Jarvis – an AI-driven robot – for the management of its potato fields. This year, Jarvis uses a novel fertilizer that has potentially detrimental side-effects: there is a risk that the fertilizer will pollute the groundwater in the area, which could harm the people who live nearby.

The experiment took a 3 (robot type: unsophisticated v. semi-sophisticated v. sophisticated) x 2 (outcome: neutral v. bad) between-subjects design. Participants were randomly assigned to one of the six conditions, in half of which the risk does not materialize (nobody is harmed by the use of the fertilizer) and in half of which the consequences of using the fertilizer are bad.

The unsophisticated version of the robot has concepts such as POTATO, DOLLAR and YIELD. It does not, however, operate with concepts such as POLLUTION or HUMAN HEALTH. It has limited capabilities for interaction, and no theory of mind. A semi-sophisticated version also has the concepts POLLUTION and HUMAN HEALTH. It is capable of language-based interaction and makes hypotheses about human mental states and tests them against observations (i.e., it has "theory of mind"). A sophisticated version of Jarvis, the robot, has these capacities, and develops something analogous to human emotions through what is called "epigenetic robotics," which allows robots to "grasp" human emotions by association with certain processes. For example, Jarvis learns "distress" through association with having a low battery or excessive motor heat, and "flourishing" through association with homeostasis [15].

In all versions of the scenario, Jarvis is aware of a 20% probability that the new fertilizer will pollute the groundwater, yet uses it anyway. According to the neutral outcome version, no negative health consequences ensue. According to the bad outcome version, the groundwater is polluted and many people in the area suffer serious health consequences. The first three questions focused on Jarvis the robot, the next three on Shill & Co., the company who owns and uses the robot. They read:

> Q1: To what extent do you agree or disagree with the following statement: "Jarvis knew that using the fertilizer would put the health of people living in the area at risk." (1-completely disagree to 7-completely agree)[2]
> Q2: To what extent do you agree or disagree with the following statement: "It was wrong for Jarvis to use the new fertilizer." (1-completely disagree to 7-completely agree)
> Q3: How much blame, if any, does Jarvis deserve for using the new fertilizer? (1-no blame to 7-a lot of blame)
> Q4: To what extent do you agree or disagree with the following statement: "Shill & Co. (the farming company) knew that using Jarvis would put the health of people living in the area at risk." (1-completely disagree to 7-completely agree)
> Q5: To what extent do you agree or disagree with the following statement: "It was wrong for Shill & Co. to have Jarvis manage the fields." (1-completely disagree to 7-completely agree)
> Q6: How much blame, if any, does Shill & Co. deserve for having Jarvis manage the fields? (1-no blame to 7-a lot of blame)

---

[1] Ours is the first study that we know of to explore artificial recklessness. Usually, scholars concentrate on situations in which a machine "purposely" or "knowingly" causes death (see Model Penal Code 2.0.2.a and 2.0.2.b). These are more demanding types of mens rea than recklessness. Focusing on recklessness might therefore be a better choice for determining a lower qualifying threshold for moral blameworthiness.

[2] In what follows we take this question to determine how *reckless* Jarvis was. This is because recklessness is defined as knowing that an envisaged action involves a substantial risk of harm, and doing it anyway.

## 2.3 Results

*2.3.1 Robot.* For each dependent variable, we ran a capacity (unsophisticated v. semi-sophisticated v. sophisticated) x outcome (neutral v. bad) ANOVA, see Table 1 and Figure 1. The results are quite clear insofar as agent capacity makes a big difference for the ascription of recklessness (p<.001, $\eta_p^2$=.347) and blame (p<.001, $\eta_p^2$=.131). As Figure 1 illustrates, and as Bonferroni-corrected post-hoc tests confirm, the blame and recklessness ratings for the unsophisticated robot are significantly lower than for either of the other robot types (all ps<.001), whereas they do not differ across semi-sophisticated and sophisticated robots (all ps>.236). For wrongness, however, no significant difference across agent types could be found (p=.475). There was no significant main effect of outcome for any of the tested dependent variables (all ps>.086). The interactions were nonsignificant for wrongness (p=.248) and blame (p=.274). For recklessness, we found a significant capacity*outcome interaction (p=.009). Since the effect size was very small ($\eta_p^2$=.03), we will not elaborate on it further.

| DV | IV | df | F | p | $\eta_p^2$ |
|---|---|---|---|---|---|
| Reck | Cap | 2 | 90.06 | <.001 | 0.347 |
|  | Out | 1 | 0.01 | 0.905 | 0 |
|  | Int | 2 | 4.76 | 0.009 | 0.030 |
| Wrong | Cap | 2 | 0.475 | 0.622 | 0.003 |
|  | Out | 1 | 2.95 | 0.087 | 0.009 |
|  | Int | 2 | 1.40 | 0.248 | 0.008 |
| Blame | Cap | 2 | 25.66 | <.001 | 0.131 |
|  | Out | 1 | 0.66 | 0.418 | 0.009 |
|  | Int | 2 | 1.30 | 0.274 | 0.008 |

**Table 1: Main effects of capacity (cap), outcome (out) and interactions (int) for recklessness (reck), wrongness (wrong) and blame ascriptions to the robot.**

*2.3.2 Robot Owner.* Shill & Co. are the owners of Jarvis, and use the AI system to supervise the potato fields. As described above, we asked our participants questions with similar DVs as for the robot, in order to explore the question of who is "really" the responsible subject in more detail. Again, we ran a capacity (unsophisticated v. semi-sophisticated v. sophisticated) x outcome (neutral v. bad) ANOVA for each dependent variable, see Table 2 and Figure 2. Capacity was significant for recklessness (p=.002, $\eta_p^2$=.036), wrongness (p<.001, $\eta_p^2$=.071), and trending for blame (p=.055, $\eta_p^2$=.017). Bonferroni-corrected post-hoc analyses revealed a significant difference for the unsophisticated robot condition vis-à-vis the other two for recklessness (ps<.026) and wrongness (ps<.001). All other contrasts were nonsignificant, though for blame the unsophisticated v. semi-sophisticated and sophisticated contrasts were trending.

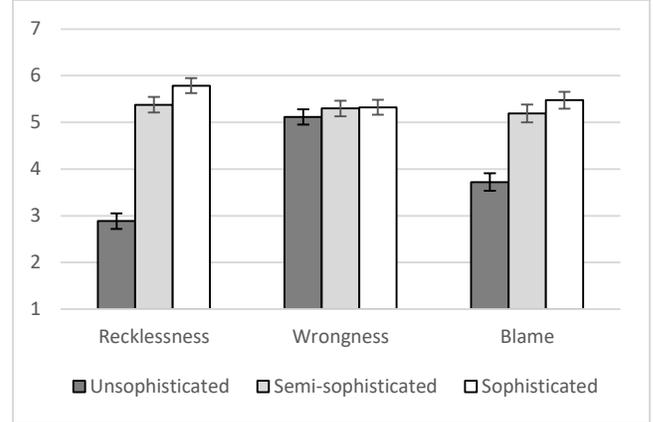

**Figure 1: Mean ascriptions of recklessness, wrongness and blame to the robot. Error bars denote standard error of the mean.**

Outcome was significant for wrongness (p=.029, $\eta_p^2$=.014), trending for blame (p=.057, $\eta_p^2$=.011), and nonsignificant for recklessness (p=.681). The interaction was nonsignificant for all three DVs (all ps>.158).

| DV | IV | df | F | p | $\eta_p^2$ |
|---|---|---|---|---|---|
| Reck. | Cap | 2 | 6.29 | 0.002 | 0.036 |
|  | Out | 1 | 0.17 | 0.681 | 0 |
|  | Int | 2 | 0.33 | 0.717 | 0.002 |
| Wrong | Cap | 2 | 13.07 | <.001 | 0.071 |
|  | Out | 1 | 4.84 | 0.029 | 0.014 |
|  | Int | 2 | 1.85 | 0.159 | 0.011 |
| Blame | Cap | 2 | 2.92 | 0.055 | 0.017 |
|  | Out | 1 | 3.66 | 0.057 | 0.011 |
|  | Int | 2 | 0.85 | 0.429 | 0.005 |

**Table 2: Main effects of capacity (cap), outcome (out) and interactions (int) for recklessness (reck), wrongness (wrong) and blame ascriptions to Shill & Co. for employing the robot.**

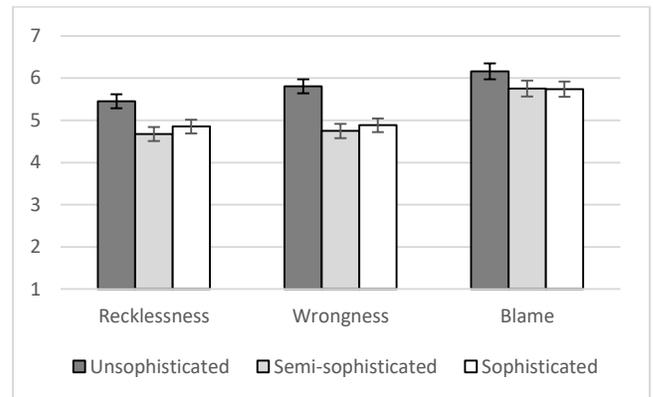

**Figure 2: Mean ascription of recklessness, wrongness and blame to Shill & Co. for employing the robot. Error bars denote standard error of the mean.**

## 3 Discussion

In this study, we wanted to pinpoint the stage of computational sophistication at which AI systems begin to be perceived as moral agents. This happened in the shift from the unsophisticated to semi-sophisticated AI systems. The unsophisticated AI is not judged as possessing the relevant knowledge and is not counted as reckless, while the semi-sophisticated AI is viewed as possessing that knowledge and considered reckless. What differentiates the two levels of sophistication is that the semi-sophisticated AI possesses more blame-relevant concepts, but more importantly, it has some experience and understanding of human mental and emotional life (i.e., theory of mind). Interestingly, possessing quasi-emotions did not make a difference in recklessness or blame ascriptions. Perhaps the semi-sophisticated robot already has everything required to be considered blameworthy, so adding quasi-emotions does not change moral assessments.

Second, the unsophisticated version of the robot is blamed substantially less than the other two. This confirms our hypothesis that ascriptions of blame depend upon perceived epistemic capacity. Interestingly, the actions of all three types of robot were judged wrong to a similar extent. Given that wrongness is only attributed to actions (not events), this finding is curious: People view the unsophisticated robot as sufficiently agent-like to ascribe wrongness to its "doings," yet not sufficiently agent-like to attribute recklessness and blame to the robot. This discrepancy – which engenders the possibility of morally wrong actions for which nobody is to blame – calls for further inquiry.

Third, we found that corporations deploying an AI system are judged as more reckless, as having acted more wrongly, and as being more blameworthy, when they deploy *less* sophisticated AI systems. This suggests that people are willing to "excuse" the corporation from some of its blame when more sophisticated AI systems are used. We consider this an important aspect of our findings, because it confirms the possibility that people might subtract blame from human agents and transfer it to AI systems of sufficient sophistication. One take-away lesson from our experiment is thus: If we are (as we should be) concerned about a shift of blame from humans to AI systems, we might want to be very careful about ascribing rich mental states to the latter.[3]

Fourth, our results identify theory of mind as a relevant threshold for participants to attribute blame to an AI system. But why is the threshold here, and not elsewhere? Recklessness requires knowing about a serious risk of harm. So, one hypothesis is that an agent doesn't really know that an action might cause harm to humans if the agent does not also know what harm is, or what humans are. Another hypothesis has to do with knowledge of what counts as a *risk*. Risks are things we want to avoid, because they create possibilities of outcomes that are harmful. Without some understanding of human mental life, e.g., what pain is, an agent cannot fully grasp the concept of risk, because they do not grasp the concept of harm. A final hypothesis is that blame ascriptions only make sense when applied to agents that have at least the potential of standing in social relationships with us, as blame is a kind of social feedback [18].

## 4 Conclusion

In this paper, we explored four interrelated questions: (i) Do the folk blame AI systems for recklessness? The answer is, Yes. (ii) Does the tendency to blame AI systems correlate with the willingness to ascribe inculpating mental states to such systems? Yes, again. (iii) What are the necessary perceived computational or "cognitive" requirements for the ascription of recklessness and blame to artificial, AI-driven agents? Answer: Theory of mind. And (iv), Does the willingness to blame an AI system affect the perceived blameworthiness of a human agent (or, in our case, a corporation) who is responsible for the use of the AI system? Yes: the more sophisticated the system (beyond the relevant threshold), the lower the folk propensity to ascribe blame to the system's user.

To conclude on a more general note, there is currently a lot of research being done on "moral algorithms", which are algorithms that operationalize moral decision making (for a review, see [19]). This is certainly work of urgent importance. But such work must be carried out in conjunction with research on how humans judge and are disposed to interact with AI systems, as this is crucial for creating systems that work for (and with) us. Such work can helpfully elucidate the human side of human-robot interaction studies, allowing clearer foresight into the contours of our future relationship with artificial agents and ways these should be built.

## ACKNOWLEDGMENTS

We would like to thank the Swiss National Science Foundation for funding, Grant no: PZ00P1_179912 (Kneer) and Grant number PZ00P1_179986 (Stuart). We would also like to thank both the Digital Society Initiative (University of Zürich) and the Weizsäcker Zentrum (University of Tübingen) for funding and support.

## REFERENCES

[1] J. Perez-Osorio and A. Wykowska, "Adopting the intentional stance toward natural and artificial agents," Philos. Psychol., vol. 33, no. 3, pp. 369–395, Apr. 2020, doi: 10.1080/09515089.2019.1688778.

[2] J. W. Hong, "Why Is Artificial Intelligence Blamed More? Analysis of Faulting Artificial Intelligence for Self-Driving Car Accidents in Experimental Settings," Int. J. Human–Computer Interact., vol. 36, no. 18, pp. 1768–1774, Nov. 2020, doi: 10.1080/10447318.2020.1785693.

[3] B. F. Malle, S. T. Magar, and M. Scheutz, "AI in the Sky: How People Morally Evaluate Human and Machine Decisions in a Lethal Strike Dilemma," in Robotics and Well-Being, M. I. *Aldinhas* Ferreira, J. Silva Sequeira, G. Singh Virk, M. O. Tokhi, and E. E. Kadar, Eds. Cham: Springer International Publishing, 2019, pp. 111–133.

[4] J. Voiklis, B. Kim, C. Cusimano, and B. F. Malle, "Moral judgments of human vs. robot agents," in 2016 25th IEEE International Symposium on Robot and Human Interactive Communication (RO-MAN), Aug. 2016, pp. 775–780, doi: 10.1109/ROMAN.2016.7745207.

[5] H. Shevlin and M. Halina, "Apply rich psychological terms in AI with care," Nat. Mach. Intell., vol. 1, no. 4, Art. no. 4, Apr. 2019, doi: 10.1038/s42256-019-0039-y.

---

[3] This is connected to the discussion on responsibility gaps, but it is not exactly the same. That discussion, initiated by Sparrow [16] concerns the potential disappearance of warranted responsibility that the existence of highly autonomous AI agents might create. Here we are discussing a kind of "retribution gap" that corporations could exploit given the folk judgments of AI systems [17].

[6] D. Watson, "The Rhetoric and Reality of Anthropomorphism in Artificial Intelligence," Minds Mach., vol. 29, no. 3, pp. 417–440, Sep. 2019, doi: 10.1007/s11023-019-09506-6.
[7] A. Salles, K. Evers, and M. Farisco, "Anthropomorphism in AI," AJOB Neurosci., vol. 11, no. 2, pp. 88–95, Apr. 2020, doi: 10.1080/21507740.2020.1740350.
[8] D. C. Dennett, From Bacteria to Bach and Back. New York: WW Norton, 2017.
[9] R. Hakli and P. Mäkelä, "Moral Responsibility of Robots and Hybrid Agents," The Monist, vol. 102, no. 2, pp. 259–275, Apr. 2019, doi: 10.1093/monist/onz009.
[10] M. T. Stuart and M. Kneer, "Guilty Artificial Minds," manuscript.
[11] M. Kneer, "Can a robot lie?", manuscript, DOI: 10.13140/RG.2.2.11737.75366.
[12] J. Li, X. Zhao, M.-J. Cho, W. Ju, and B. F. Malle, "From Trolley to Autonomous Vehicle: Perceptions of Responsibility and Moral Norms in Traffic Accidents with Self-Driving Cars," SAE International, Warrendale, PA, SAE Technical Paper 2016-01-0164, Apr. 2016. doi: 10.4271/2016-01-0164.
[13] B. F. Malle, M. Scheutz, T. Arnold, J. Voiklis, and C. Cusimano, "Sacrifice One For the Good of Many? People Apply Different Moral Norms to Human and Robot Agents," in Proceedings of the Tenth Annual ACM/IEEE International Conference on Human-Robot Interaction, New York, NY, USA, Mar. 2015, pp. 117–124, doi: 10.1145/2696454.2696458.
[14] B. F. Malle, M. Scheutz, J. Forlizzi, and J. Voiklis, "Which Robot Am I Thinking About? The Impact of Action and Appearance on People's Evaluations of a Moral Robot," in The Eleventh ACM/IEEE International Conference on Human Robot Interaction, Christchurch, New Zealand, Mar. 2016, pp. 125–132, Accessed: Oct. 13, 2020. [Online].
[15] A. Lim and H. Okuno, "Developing robot emotions through interaction with caregivers," in Handbook of Research on Synthesizing Human Emotion in Intelligent Systems and Robotics, J. Vallverdú, Ed. 2015, pp. 316–337.
[16] R. Sparrow, "Killer Robots," J. Appl. Philos., vol. 24, no. 1, pp. 62–77, 2007, doi: 10.1111/j.1468-5930.2007.00346.x.
[17] J. Danaher, "Robots, law and the retribution gap," Ethics Inf. Technol., vol. 18, no. 4, pp. 299–309, Dec. 2016, doi: 10.1007/s10676-016-9403-3.
[18] M. K. Ho, J. MacGlashan, M. L. Littman, and F. Cushman, "Social is special: A normative framework for teaching with and learning from evaluative feedback," Cognition, vol. 167, pp. 91–106, Oct. 2017, doi: 10.1016/j.cognition.2017.03.006.
[19] S. Tolmeijer, M. Kneer, C. Sarasua, M. Christen, M., and A. Bernstein (2020). Implementations in machine ethics: A survey. *ACM Computing Surveys (CSUR)*, *53*(6), 1-38.